\documentclass[cits]{PoS}
\newcommand{\MET}{\slash\!\!\!\!E_T}

\title{Trilepton production at the CERN LHC: SUSY Signals and Standard Model Backgrounds}
\ShortTitle{SUSY Signals and Standard Model Backgrounds}

\author{\speaker{Edmond BERGER}%
         \thanks{Argonne report number ANL-HEP-CP-09-92.  Research supported by the 
         U.~S.\ Department of Energy under Contract No.~DE-AC02-06CH11357.  We gratefully 	acknowledge the use of JAZZ, a 350-node
computer cluster operated by the Mathematics and Computer Science Division at
Argonne as part of the Laboratory Computing Resource Center.}\\
        Argonne National Laboratory\\
        E-mail: \email{berger@anl.gov}}

\author{Zack SULLIVAN\\
        Illinois Institute of Technology\\
        E-mail: \email{Zack.Sullivan@iit.edu}}

\abstract{Events with isolated leptons and missing energy in the final state are known 
to be signatures of new physics phenomena at high energy collider physics
facilities.   Standard model (SM) sources of isolated trilepton final states include
gauge boson pair production such as $WZ$ and $W\gamma^{*}$, and $t\bar t$
production.  Symbol $\gamma^*$ represents a virtual photon.  
Our new contribution is the demonstration that bottom and charm meson decays, $b
\rightarrow l X$ and $c \rightarrow l X$, produce isolated lepton
events that can overwhelm the effects of other processes.  We compute
contributions from a wide range of SM heavy flavor processes including $b
Z/{\gamma^{*}}$, $c Z/ \gamma^{*}$, $b \bar{b} Z/ \gamma^{*}$, $c \bar{c} Z/
\gamma^{*}$.  We also include contributions from processes in which a $W$ is 
produced in association with one or more
heavy flavors such as $t W$, $b \bar{b} W$, $c \bar{c} W$.  In all these
cases, one or more of the final observed isolated leptons comes from a heavy
flavor decay.   We propose new cuts to control the heavy flavor 
backgrounds in the specific case of chargino plus neutralino pair production
in supersymmetric models.}


\FullConference{European Physical Society Europhysics Conference on High Energy Physics,
EPS-HEP 2009,\\
		 July 16 - 22 2009\\
		 Krakow, Poland}

\begin{document}

\section{Introduction}
{\em Isolated} leptons along with missing transverse energy $\MET$ are signatures for new physics processes at collider energies.  A known example of charged dilepton production is Higgs boson decay,  $H \rightarrow W^+ W^-$ followed by purely leptonic decay of the $W$ intermediate vector bosons.  Charged trilepton production may arise from the associated production of a chargino $\tilde {\chi}_1^{\pm}$ and a neutralino $\tilde{\chi}^0_2$ in supersymmetric models, followed by the leptonic decays of the chargino and neutralino.   There are many standard model (SM) sources of isolated leptons, such as leptonic decays of $W$ and $Z$ bosons produced from standard model processes.  Semi-leptonic decays of heavy flavors (bottom and charm quarks) also make a very important contribution to the rate of isolated lepton production.  The nature and magnitude of these contributions from heavy flavor sources are emphasized in our two recent papers~\cite{Sullivan:2006hb,Sullivan:2008ki} and summarized in this brief report.  

\section{Isolated leptons from heavy flavor decays}

Given a lepton track and a cone, in rapidity and azimuthal angle space, of size $\Delta R$, the lepton is said to be {\em isolated} if the sum of the transverse energy of all other particles within the cone is less than a predetermined value (either a constant or a value that scales with the transverse momentum of the lepton).  Our simulations based on the known semi-leptonic decays of bottom and charm mesons show that leptons which satisfy isolation take a substantial fraction of the momentum of the parent heavy meson.  Moreover, isolation leaves $\sim 7.5 \times 10^{-3}$ muons per parent $b$ quark.  The potential magnitude of the background from heavy flavor decays may be appreciated from the fact that one 
begins with an inclusive $b \bar{b}$ cross section at LHC energies of about $5 \times 10^8$~pb.  A suppression of $\sim 10^{-5}$ from isolation still leaves a formidable rate of isolated dileptons.  For the isolated leptons, our simulations show that roughly $1/2$ of the events satisfy isolation because the remnant is just outside whatever cone is used for the tracking and energy cuts, and another $1/2$ pass because the lepton took nearly all the energy, meaning there is nothing left to reject upon.  The latter events are not candidates to reject with impact parameter cuts since they tend to point to the primary vertex.  Although the decay leptons are ``relatively'' soft, we find that their associated backgrounds extend well into the region of new physics with relatively large mass scales, such as a Higgs boson with mass $\sim 160$~GeV.  

\section{SM backgrounds in Higgs boson production and decay}

Our analysis of the role of heavy flavor backgrounds in  $H \rightarrow W^+ W^-\rightarrow l^+ l^- + \MET$ at Fermilab Tevatron and CERN Large Hadron Collider (LHC) energies is presented in Ref.~\cite{Sullivan:2006hb}.  In addition to continuum $W^+ W^-$, $Z/\gamma^*$, and $t \bar{t}$, we simulate the contributions from processes with $b$ and $c$ quarks in the final state, including $b \bar{b} X$, $c \bar{c} X$, $W c$, $W b$, $W b \bar{b}$, as well as single top quark contributions.  Symbol $\gamma^*$ represents a virtual photon (a ``Drell-Yan'' pair of leptons).  We use QCD hard matrix elements fed through PYTHIA showering.  The PYTHIA output is then put through a detector simulation code.  We learn that isolation cuts do not generally remove leptons from heavy flavor sources as backgrounds to multi-lepton searches.  A sequence of complex physics cuts is needed, conditioned by the new physics one is searching for.  Moreover, the heavy flavor backgrounds cannot be easily extrapolated from more general samples.  The interplay between isolation and various physics cuts tends to emphasize corners of phase space rather than the bulk characteristics.  Nevertheless, for Higgs boson searches in the mass range $\sim 160$~GeV, we find that hardening the cut on the momentum of the next-to-leading lepton serves to suppress heavy flavor backgrounds adequately at LHC energies~\cite{Sullivan:2006hb}.   

\section{Trileptons at the LHC}   
The associated production of a chargino and neutralino, followed by their leptonic decays, 
 $\tilde {\chi}_1^{\pm} \tilde{\chi}^0_2 \rightarrow l^+ l^- l^{\pm} + \MET$ is a golden signature for supersymmetry.  The LHC collaborations ATLAS and CMS have devised strategies to observe this signal, as reported in their respective Technical Design Reports 
 (TDRs)~\cite{Aad:2009wy, Ball:2007zza}.   The SM backgrounds examined in detail include continuum $W Z$ and $W \gamma^*$ production and leptonic decay, along with 
 $t \bar {t}$, $t W$, and $t \bar{b}$ production and decay.    In Ref.~\cite{Sullivan:2008ki}, we repeat the CMS and ATLAS simulations of the SUSY signals and SM backgrounds, but we include, in addition, the contributions to the backgrounds from $b Z/\gamma^*$, $b \bar {b} Z/\gamma^*$, $c Z/\gamma^*$, $c \bar {c} Z/\gamma^*$, $b \bar {b} W$, and $c \bar {c} W$.  To touch base with the CMS and ATLAS analyses, we examine the SUSY trilepton signal and SM backgrounds for four SUSY points labeled LM1, LM7, LM9, and SU2.   Their parameter values may be found in Ref.~\cite{Sullivan:2008ki}.  These points may be disfavored by other data, but we adopt them to make contact with the ATLAS and CMS simulations.  
 
We reproduce the analysis chains described in Refs.~\cite{Aad:2009wy, Ball:2007zza}.  Our hard-scattering matrix elements are computed with MadEvent~\cite{Maltoni:2002qb} at leading-order (LO) in perturbation theory, so that we retain all spin and angular correlations.  We feed the LO results into PYTHIA in order to include the effects of showering and hadronization.  The LO treatment is perhaps adequate in view of the rejection for physics reasons of events with hard jets, and because we want to avoid double-counting of radiation included in PYTHIA.  An alternative approach would begin with next-to-leading (NLO) order matrix elements and a showering code that deals properly with matching and double counting aspects of the radiation.  Not having this 
tool available, and recognizing that any showering code will have its limitations until it has been tested and tuned against LHC data, 
we proceed as described.  Our MadEvent results, fed through PYTHIA showering and then through a detector simulation, reproduce the CMS and ATLAS full detector results to 10\%.  
The important cuts in the physics analysis are (a) a requirement of 3 isolated leptons with transverse momenta $p_{T, \mu} > 10$~GeV, $p_{T, e} > 17$~GeV; (b) a requirement that there be no jets with $E_T > 30$~GeV, to reduce effects from $t \bar{t}$ production and from higher mass SUSY  sources; and (c) a requirement that the invariant mass of a pair of opposite-sign, same-flavor (OSSF) leptons
$M_{ll}^{OSSF} < 75$~GeV to eliminate backgrounds from real $Z$ bosons.  As is 
detailed in Ref.~\cite{Sullivan:2008ki}
the contributions of $Z/\gamma^*$ plus heavy flavor decays produce trileptons 10 times more often than the previously considered 
$W Z/\gamma^*$ source in the region below the $Z$ peak.  The SUSY signals are overwhelmed.    

The number of additional cuts available to reject the background from $Z/\gamma^*+$heavy flavors is 
limited.  In Ref.~\cite{Sullivan:2006hb} we recommend raising the minimum lepton $p_T$ threshold 
since the lepton $p_T$ spectrum from $b$ and $c$ decays tends to fall rapidly.  In typical trilepton
studies, however, the leptons are soft, and an increase in the cut on the lepton $p_T$ tends to reject 
too much of the signal.  Missing transverse energy $\MET$ is somewhat discriminatory.  The SUSY signals
contain invisible neutralinos which leave a broad range of $\MET$ in the
detector.  Trilepton signatures from $t\bar t$ production generally have two neutrinos which lead
to large missing energy.  The contribution from $Z/\gamma^*+$heavy flavor
processes peaks at over 400 times the size of the LM9 signal at low $\MET$,
but it falls rapidly to below the signal by $\MET>50$ GeV.  

We find that the requirement $\MET>30$ GeV removes a reasonable fraction of the $Z/\gamma^*+$heavy flavor backgrounds for a modest loss of signal.  A cut below 20 GeV is not as useful and is likely not achievable at the LHC.  A cut above 40 GeV removes most of the $Z/\gamma^*+X$
backgrounds, but it begins to significantly reduce the signal and is of little
additional help with $WZ/\gamma^*$ and $t\bar t$ backgrounds.  The sharply
falling $\MET$ spectrum in $Z/\gamma^*+X$ is sensitive to
uncertainties in the measurement of $\MET$.  This uncertainty makes it
difficult to predict absolute cross sections after cuts.  On the other hand,
this sensitivity could provide an opportunity to \textit{measure the
background in situ} and reduce concerns regarding modeling details.  The
background can be fit in the data and the $\MET$ cut adjusted to optimize the
purity of the sample.  

Since the accuracy of $\MET$ measurements is limited, we examine also the
utility of angular cuts.  There are significant angular
correlations in the $Z/\gamma^*+$heavy flavor backgrounds that are different
from those in the SUSY trilepton signals or the $WZ/\gamma^*$ and $t\bar t$
backgrounds.  We examine the angular
distribution $\theta_{ij}^{\mathrm{CM}}$ between pairs of $p_T$-ordered
leptons in the trilepton center-of-momentum (CM) frame.
The $Z/\gamma^*+$heavy flavor backgrounds have significant peaks at both small
and large angles.  The signal and other backgrounds either peak only at large
angles, or are fairly central.

\section{Summary}
We find that the dominant backgrounds to low-momentum trilepton
signatures come from real $b$ and $c$ decays.  For the CMS and ATLAS
SUSY analyses we examine, the $Z/\gamma^*+$heavy flavor decay
backgrounds are a factor of 10--30 larger than $WZ/\gamma^*$ or $t\bar{t}$ to
trileptons.  Large $\MET$ cuts and angular correlations can be used to
significantly reduce the heavy flavor backgrounds, but we must be mindful of
the modest $\MET$ in the SUSY signal.  Along with our results for dileptons
in Ref.~\cite{Sullivan:2006hb}, we argue that leptons from heavy flavor
decays should be examined for all low-momentum lepton signals.  Once
normalizations are measured with LHC data, we may have handles to reduce the
effect of these backgrounds to an acceptable level.  The overall message is that precise 
understanding of all SM physics processes will enable confident discovery claims.

\end{document}